\documentclass{article}
\usepackage{spconf,amsmath,epsfig}

\title{Super-resolved multi-temporal segmentation with deep permutation-invariant networks}
\name{Diego Valsesia, Enrico Magli}
\address{Politecnico di Torino}

\begin{document}
%
\maketitle
\begin{abstract}
Multi-image super-resolution from multi-temporal satellite acquisitions of a scene has recently enjoyed great success thanks to new deep learning models. In this paper, we go beyond classic image reconstruction at a higher resolution by studying a super-resolved inference problem, namely semantic segmentation at a spatial resolution higher than the one of sensing platform. We expand upon recently proposed models exploiting temporal permutation invariance with a multi-resolution fusion module able to infer the rich semantic information needed by the segmentation task. The model presented in this paper has recently won the AI4EO challenge on Enhanced Sentinel 2 Agriculture.  
\end{abstract}
\section{Introduction}
\label{sec:intro}
Image super-resolution is the process of recovering information from one or more degraded images, sampled at a lower spatial resolution (LR). This process is typically ill-posed so strong priors on the data are needed to effectively address the problem and, for this reason, deep learning has shown impressive results in recent years. Multi-temporal image super-resolution (MISR) \cite{1331445,KATO201564} is particularly interesting in the context of remote sensing, since multiple versions of the same scene are naturally acquired by revisits of a satellite. Thanks to sub-pixel phase shifts, multiple images carry complementary information about the scene and impressive super-resolution results can be achieved with suitable fusion techniques. Very recent works have developed deep-learning models that can combine multiple images in complex non-linear fashions to reconstruct the higher-resolution version of a scene. In particular, DeepSUM \cite{molini2019deepsum} proposed a modular architecture and the idea of adaptive filters that are computed directly from the features extracted by the neural network to register and interpolate the multiple low-resolution images. Highres-net \cite{rarefin2020multi} developed a recursive fusion approach to the problem. RAMS \cite{salvetti2020multi} improved the state of the art by careful usage of the feature attention mechanism. Most recently, PIUnet \cite{valsesia2021permutation} showed that temporal ordering of the input images does not matter and designing an architecture that is fully invariant to temporal permutation captures richer representations, improving reconstruction performance and data efficiency.

The aforementioned methods and most of the literature have been focused on image generation, whereby the output of the model is the image itself, sampled at a higher spatial resolution. On the other hand, in this paper, we study a problem of super-resolved inference, i.e., inferring some property with increased spatial resolution, which is a problem that has been largely neglected by MISR models in remote sensing. In particular, we work in the context of the AI4EO challenge on Enhanced Sentinel 2 Agriculture \cite{ai4eo}, where multi-temporal Sentinel 2 images are used for the detection of cultivated areas. However, the segmentation map of cultivated areas must be produced with a resolution of 2.5 meters per pixel, while the input Sentinel 2 bands have resolutions up to 10 meters. Thus, an effective model for the problem needs to suitably combine the multiple input images to extract spatially super-resolved features which, at the same time, are also semantically discriminative for the detection of cultivated areas.

\section{Proposed method}
\label{sec:method}
\begin{figure*}
    \centering
    \includegraphics[width=\textwidth]{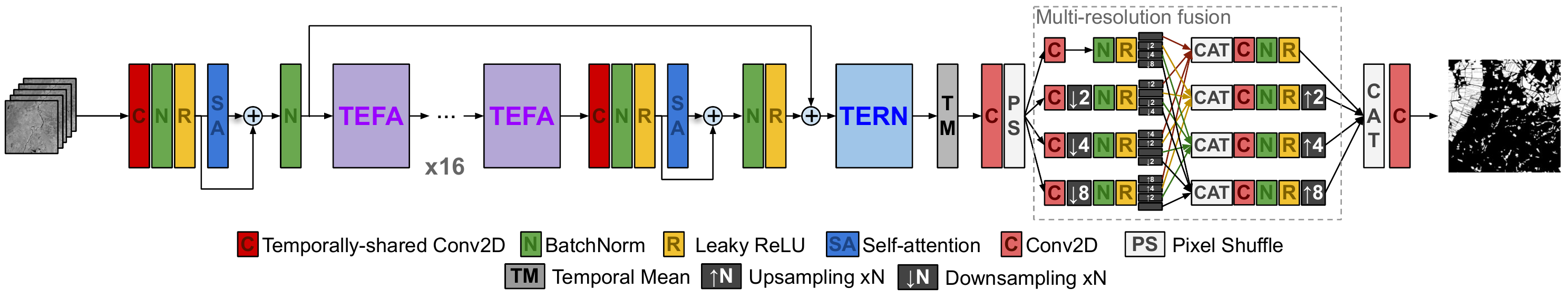}
    \vspace*{-6pt}
    \caption{SPInet architecture.}
    \vspace*{-6pt}
    \label{fig:architecture}
\end{figure*}

\begin{figure}
    \centering
    \includegraphics[width=0.8\columnwidth]{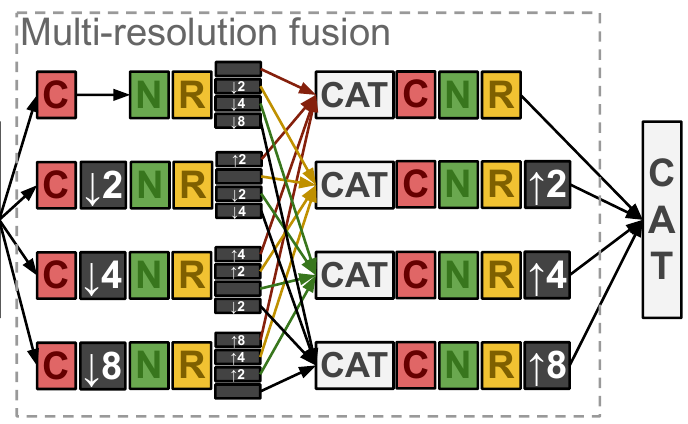}
    \vspace*{-6pt}
    \caption{Multi-resolution fusion block. Features are extracted at four different scales and scales are made to densely interact with other before returning features at the highest scale only. This promotes more semantically meaningful features, requiring larger receptive field and multi-resolution reasoning.}
    \vspace*{-6pt}
    \label{fig:mf_block}
\end{figure}

In this section, we present an overview of the proposed model to address super-resolved segmentation of cultivated areas from Sentinel 2 images. The model is called SPInet (Segmentation by Permutation Invariance) and an overview is shown in Fig.\ref{fig:architecture}. SPInet treats the super-resolved segmentation problem in an end-to-end fashion, i.e., it is trained to directly regress the super-resolved segmentation maps, rather than generating super-resolved images and then applying a segmentation model on those. The reason for this design is efficiency, since a single end-to-end model can use its parameters in the most efficient manner, reducing computational and data requirements.

Existing models for MISR have largely neglected the issue of temporal permutation, whereby the temporal ordering of the input images does not carry any relevant information for the super-resolution task and causes such models to be inefficient with the, often scarce, ground truth data available for training. Valsesia et al. \cite{valsesia2021permutation} have recently shown how models ought not to learn feature extractors that rely on temporal ordering, but rather should be invariant to temporal permutation by design. In this work, we build a model that is fully invariant to temporal permutation for the task of image segmentation at an increased spatial resolution. The backbone of the model is the PIUnet architecture \cite{valsesia2021permutation}, but the segmentation problem significantly differs from image generation. While super-resolved image generation largely exploits localized features, segmentation requires larger receptive fields and interaction between features across scales in order to infer semantic properties.  

Following Fig.\ref{fig:architecture}, the first part of the architecture up to the temporal mean operation closely follows PIUnet, with the exception of the first convolutional layer which processes multiple bands instead of a single one. One of the main design patterns is the use of a temporally-shared 2D convolution operation followed by batch normalization, leaky ReLU non-linearity and a residual self-attention operation \cite{vaswani2017attention} in the temporal dimension. This block is, by design, equivariant to temporal permutation and serves as a fundamental building block to extract spatio-temporal features. The Temporally-Equivariant Feature Attention (TEFA) block uses this fundamental building block to construct a variant of classic feature attention that is equivariant to temporal permutations. The repetition of 16 TEFA blocks serves as the backbone of our network. The Temporally-Equivariant RegNet (TERN) is a module that proposes and applies adaptive registration filters to the features and allows to compensate for misregistration, always maintaining temporal equivariance. The temporal mean operation, being symmetric over the temporal dimension, allows to transform the model from equivariant to invariant. Notice how there is no invariance in the spectral dimension because ordering does matter when processing spectral information. 

The promotion of semantically meaningful features is performed with the inclusion of a multi-resolution feature fusion block (see Fig.\ref{fig:mf_block}). This serves as an enhancement for the segmentation task, with respect to pure super-resolution, because it allows features to exploit a more global context, rather than localized properties. This block is inspired by state-of-the-art operations used in semantic segmentation networks such as HRNet \cite{wang2020deep}. It generates four dyadic scales where features at different resolution and spatial receptive fields can be extracted before merging them to create the output logits. More in detail, after the pixel shuffling upsampling operation, four strided convolutions create four parallel branches whose spatial resolution is either the original (i.e. the super-resolved one), $1/2$, $1/4$, or $1/8$. Each branch is further downsampled or upsampled to generate versions of itself at all the four scales. Then, features from highest resolution from all branches are concatenated into a new high-resolution branch. Similarly, the features from all versions at $1/2$ resolution are concatenated into a $1/2$ resolution branch and so on for all the scales. After a convolution on each branch, the lower scales are upsampled and the features from all branches are concatenated and projected to generate the segmentation map.

Finally, we remark that the model does not require a fixed number of input images and can work with any number of low-resolution images, depending on availability and quality, without retraining or redesigning the architecture. Expensive temporal self-ensembling operations where the model performance is boosted by averaging the results of different input temporal permutations are avoided thanks to the invariance property which already captures this richness in the model.

\section{Experimental results}
\label{sec:results}

\subsection{Experimental setting}

We tested the proposed model on the dataset provided by the AI4EO challenge on Enhanced Sentinel 2 Agriculture \cite{ai4eo}. The public dataset is composed of 100 scenes, each with a variable number of multi-temporal images (usually more than 15) collected between March 1st and September 1st 2019. All the Sentinel 2 bands are either native or interpolated to 10m per pixel and they are Bottom-of-Atmosphere L2A products with $500 \times 500$ pixels. Ground truth segmentation maps are provided with size $2000\times2000$ at 2.5m per pixel. We partitioned the public data into training and validation sets, with 90 and 10 scenes each. For training purposes, a fixed number of multi-temporal images, equal to 20, is used. Cloudless images are selected whenever possible. Testing used all available multi-temporal images with at most $25\%$ cloud coverage. All the 12 bands of Sentinel 2 have been used in this work, including the 20m and 60m ones, which have been interpolated to 10m.

SPInet has been trained for about 60 epochs with a batch size equal to 8 and learning rate equal to $10^{-4}$ with the binary cross-entropy loss function. The number of feature channels is set to 48 throughout the network. Feature attention uses a bottleneck with 6 features. Patches of size $32\times32$ have been extracted for the training process. Model training required about two days with a Titan RTX GPU with peak VRAM occupation equal to 16GB.

\subsection{Segmentation results}
We compare the performance of SPInet on our validation set against two relevant baselines. The first baseline is a model-based segmentation approach which relies on bicubic interpolation and thresholding of the temporally-averaged normalized difference vegetation index (NDVI). Additionally, we include the results of a deep learning baseline consisting in using a state-of-the-art neural network for semantic segmentation, namely DeepLabv3 \cite{chen2017rethinking} applied on the bicubically-interpolated and temporally-averaged images.

Performance is evaluated in terms of the Matthews correlation coefficient (MCC) which better accounts for  true and false positives (TP,FP) and negatives (TN,FN) with respect to other measures like F1 score or accuracy, especially with unbalanced classes. The MCC is defined as:
\begin{align*}
    \text{MCC} = \frac{TP \times TN - FP \times FN}{\sqrt{(TP+FP)(TP+FN)(TN+FP)(TN+FN)}}.
\end{align*}

Table \ref{table:mcc} shows the MCC obtained by the different methods on the validation set. It can be noticed that the proposed end-to-end approach with SPInet significantly outperforms the simple model-based baseline and also improves upon the deep-learning segmentation approach.

\begin{table}
\centering
\caption{Cultivated land detection - MCC}
\label{table:mcc}
\begin{tabular}{lcccc}
& Model-based & Deeplabv3 & \textbf{SPInet} \\ \hline \hline
\text{eopatch-902} & 0.636 & 0.785 & \textbf{0.895} \\
\text{eopatch-903} & 0.413 & 0.704 & \textbf{0.802} \\
\text{eopatch-904} & 0.550 & 0.721 & \textbf{0.819} \\
\text{eopatch-905} & 0.604 & 0.804 & \textbf{0.895} \\
\text{eopatch-907} & 0.534 & 0.692 & \textbf{0.802} \\
\text{eopatch-908} & 0.544 & 0.702 & \textbf{0.787} \\
\text{eopatch-914} & 0.474 & 0.652 & \textbf{0.778} \\
\text{eopatch-915} & 0.325 & 0.507 & \textbf{0.689} \\
\text{eopatch-916} & 0.525 & 0.659 & \textbf{0.819} \\
\text{eopatch-924} & 0.496 & 0.605 & \textbf{0.736} \\
\hline
Avg. MCC & 0.510 & 0.683 & \textbf{0.802}\\ \hline
\end{tabular}%
\end{table}

\begin{table}
\centering
\caption{Effectiveness of multi-resolution fusion (MRF) - MCC}
\label{table:nomrf}
\begin{tabular}{lcccc}
& \textbf{SPInet with MRF} & SPInet w/o MRF \\ \hline \hline
Avg. MCC & \textbf{0.802}  & 0.783 \\ \hline
\end{tabular}%
\end{table}

\begin{figure*}
    \centering
    \includegraphics[width=0.195\textwidth]{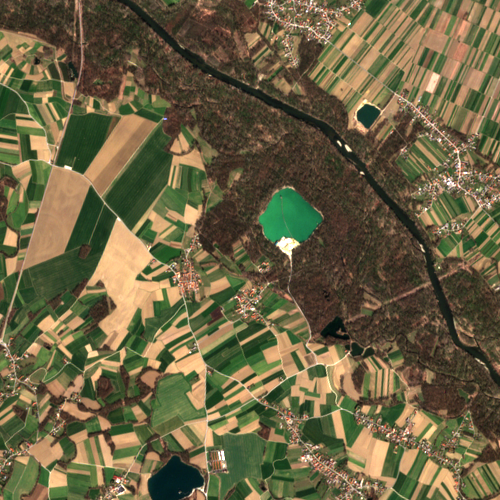}
    \includegraphics[width=0.195\textwidth]{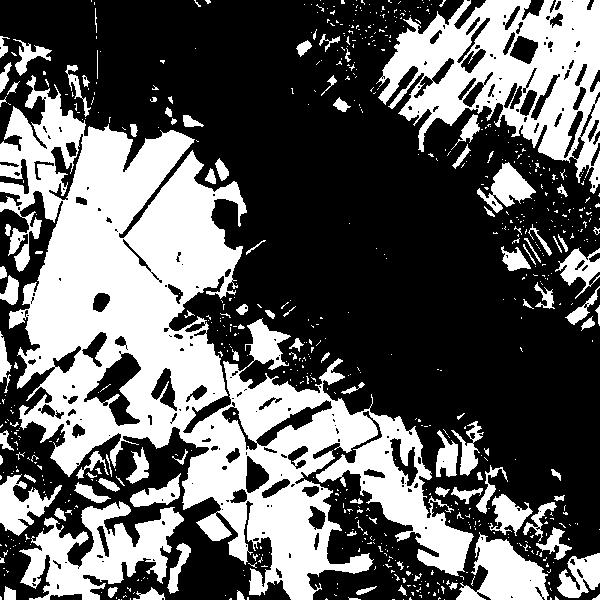}
    \includegraphics[width=0.195\textwidth]{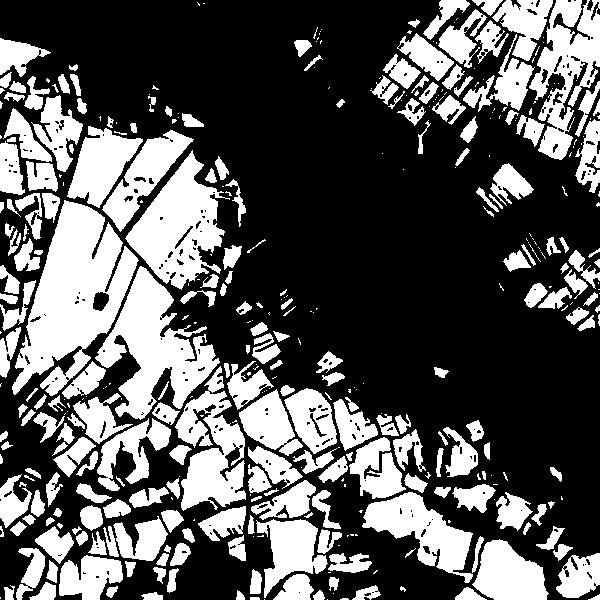}
    \includegraphics[width=0.195\textwidth]{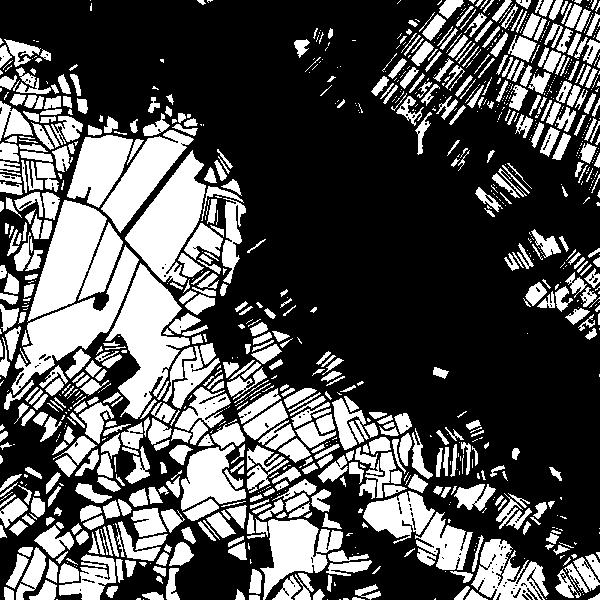}
    \includegraphics[width=0.195\textwidth]{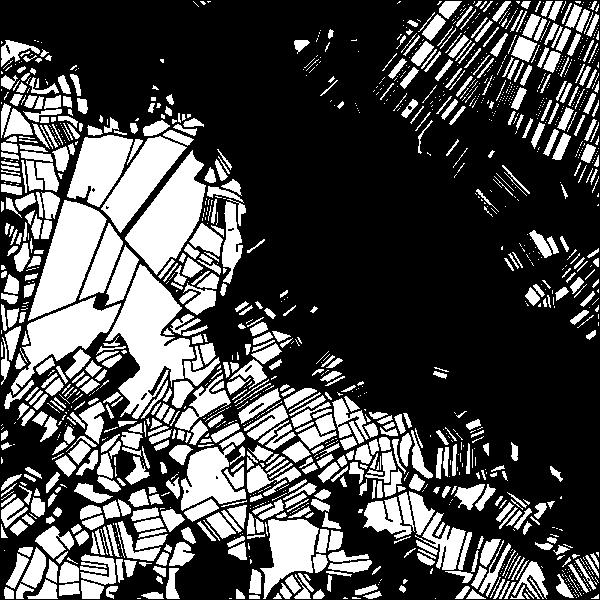}
    \caption{Results on \textit{eopatch-902}. Left to right: one of the LR images, cultivated areas map produced by model-based approach, bicubic+Deeplab deep learning approach, SPInet and ground truth.}
    \label{fig:val1}
\end{figure*}

\begin{figure*}
    \centering
    \includegraphics[width=0.195\textwidth]{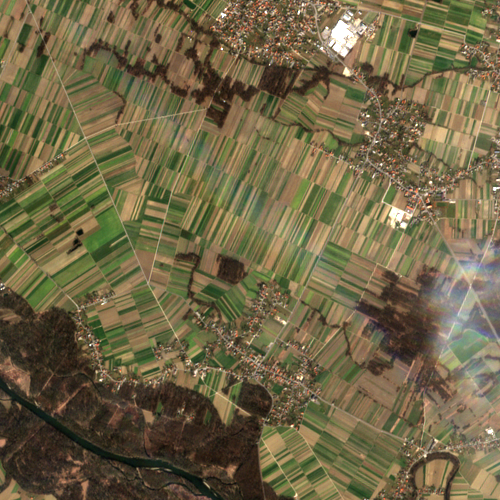}
    \includegraphics[width=0.195\textwidth]{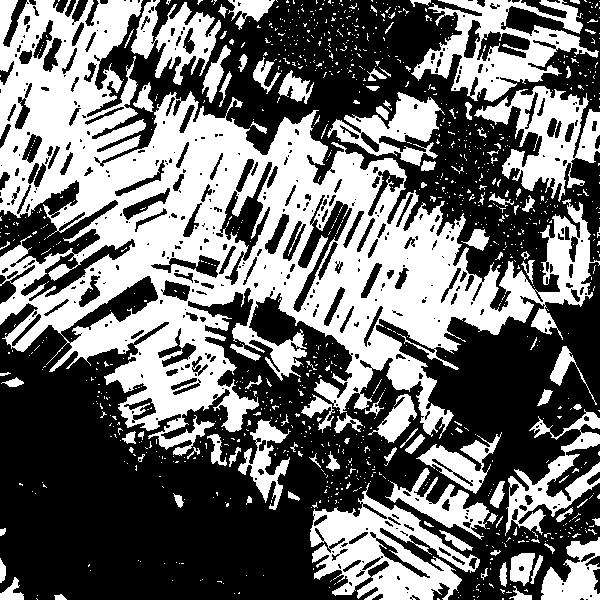}
    \includegraphics[width=0.195\textwidth]{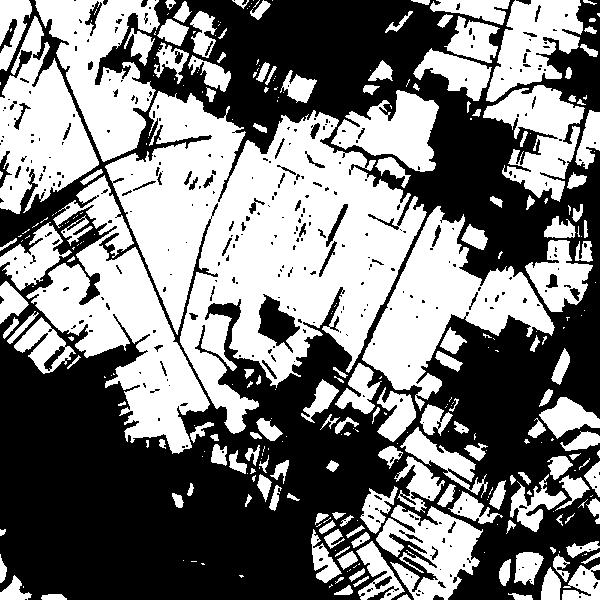}
    \includegraphics[width=0.195\textwidth]{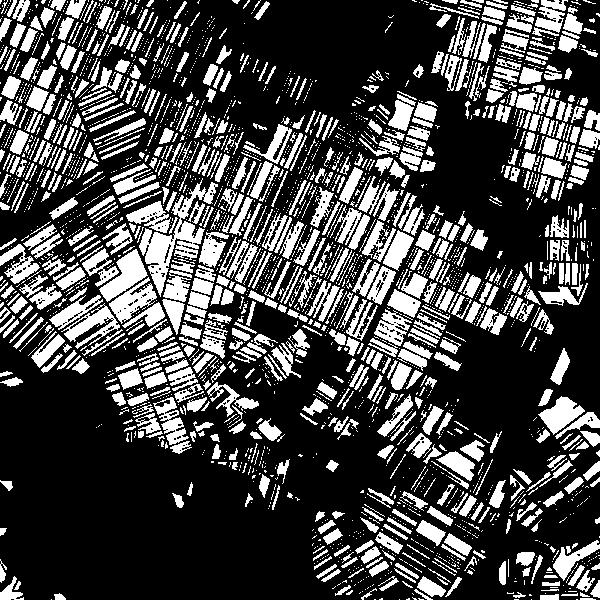}
    \includegraphics[width=0.195\textwidth]{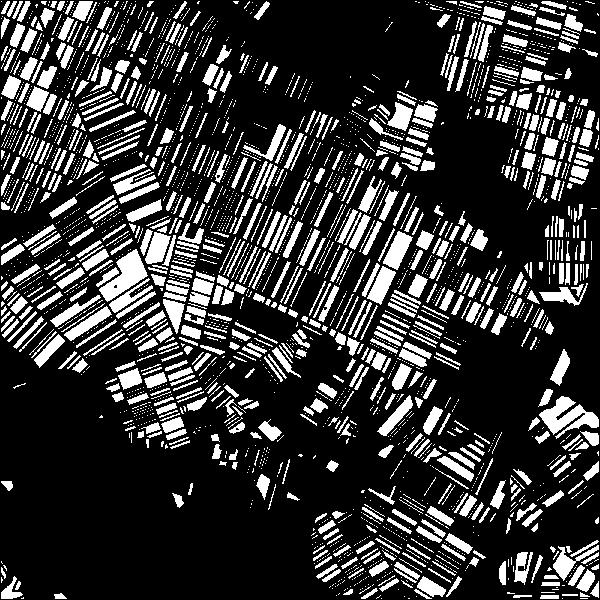}
    \caption{Results on \textit{eopatch-914}. Left to right: one of the LR images, cultivated areas map produced by model-based approach, bicubic+Deeplab deep learning approach, SPInet and ground truth.}
    \label{fig:val2}
\end{figure*}

Figures \ref{fig:val1} and \ref{fig:val2} show the segmentation results obtained by the various methods on two validation images when compared against the ground truth segmentation map. It can be noticed how the proposed method is able to capture finer details of the cultivated areas map, showing its effectiveness in generating a super-resolved result. Existing segmentation techniques such as the tested DeepLab model heavily rely on downsampling because the classic semantic segmentation tasks studied in computer vision do not carry much fine spatial information. This is undesirable for the super-resolved cultivated land detection problem we study and the proposed joint super-resolution and segmentation approach avoids such operations and results in better performance.

Table \ref{table:nomrf} reports an ablation study that highlights the importance of the multi-resolution fusion module. In this experiment, the multi-resolution fusion module is replaced by two convolutional layers interleaved with batch normalization and ReLU maintaining the full image resolution and with overall a comparable number of weights with respect to the fusion block. This solution appears to be suboptimal, as expected, because there is no mechanism to exploit multi-resolution representations which are important to capture scene semantics.

Finally, as a reference, we report that SPInet has achieved an MCC value of 0.862 on the private test set of the AI4EO challenge.

\section{Conclusions}
\label{sec:conclusions}
In this paper, we presented an end-to-end deep learning approach for a super-resolved inference problem consisting in detecting cultivated areas with higher spatial resolution than what is offered by the input Sentinel 2 images. We leveraged recent results on the importance of temporal permutation invariance in the design of deep neural network that deal with multi-temporal data, and specifically for MISR. We also showed how multi-resolution fusion is important in this inference problem with respect to pure image super-resolution. Future work will focus on estimating and exploiting aleatoric uncertainty which was shown in \cite{valsesia2021permutation} to be an important factor in presence of temporal variation in the input data.

\bibliographystyle{IEEEbib}
\bibliography{biblio}

\begin{thebibliography}{10}

\bibitem{1331445}
S.~{Farsiu}, M.~D. {Robinson}, M.~{Elad}, and P.~{Milanfar},
\newblock ``Fast and robust multiframe super resolution,''
\newblock {\em IEEE Transactions on Image Processing (TIP)}, vol. 13, no. 10,
  pp. 1327--1344, Oct 2004.

\bibitem{KATO201564}
Kato Toshiyuki, Hino Hideitsu, and Murata Noboru,
\newblock ``Multi-frame image super resolution based on sparse coding,''
\newblock {\em Neural Networks}, vol. 66, pp. 64 -- 78, 2015.

\bibitem{molini2019deepsum}
Andrea~Bordone Molini, Diego Valsesia, Giulia Fracastoro, and Enrico Magli,
\newblock ``Deepsum: Deep neural network for super-resolution of unregistered
  multitemporal images,''
\newblock {\em arXiv preprint arXiv:1907.06490}, 2019.

\bibitem{rarefin2020multi}
Md~Rifat Arefin, Vincent Michalski, Pierre{-}Luc St{-}Charles, Alfredo
  Kalaitzis, Sookyung Kim, Samira~Ebrahimi Kahou, and Yoshua Bengio,
\newblock ``Multi-image super-resolution for remote sensing using deep
  recurrent networks,''
\newblock in {\em 2020 {IEEE/CVF} Conference on Computer Vision and Pattern
  Recognition, {CVPR} Workshops 2020, Seattle, WA, USA, June 14-19, 2020}.
  2020, pp. 816--825, {IEEE}.

\bibitem{salvetti2020multi}
Francesco Salvetti, Vittorio Mazzia, Aleem Khaliq, and Marcello Chiaberge,
\newblock ``Multi-image super resolution of remotely sensed images using
  residual attention deep neural networks,''
\newblock {\em Remote Sensing}, vol. 12, no. 14, 2020.

\bibitem{valsesia2021permutation}
Diego Valsesia and Enrico Magli,
\newblock ``Permutation invariance and uncertainty in multitemporal image
  super-resolution,''
\newblock {\em arXiv preprint arXiv:2105.12409}, 2021.

\bibitem{ai4eo}
``{AI4EO Enhanced Sentinel 2 Agriculture},''
  \url{https://platform.ai4eo.eu/enhanced-sentinel2-agriculture},
\newblock Accessed: 2010-10-13.

\bibitem{vaswani2017attention}
Ashish Vaswani, Noam Shazeer, Niki Parmar, Jakob Uszkoreit, Llion Jones,
  Aidan~N Gomez, {\L}ukasz Kaiser, and Illia Polosukhin,
\newblock ``Attention is all you need,''
\newblock in {\em Proceedings of the 31st International Conference on Neural
  Information Processing Systems}, 2017, pp. 6000--6010.

\bibitem{wang2020deep}
Jingdong Wang, Ke~Sun, Tianheng Cheng, Borui Jiang, Chaorui Deng, Yang Zhao,
  Dong Liu, Yadong Mu, Mingkui Tan, Xinggang Wang, et~al.,
\newblock ``Deep high-resolution representation learning for visual
  recognition,''
\newblock {\em IEEE transactions on pattern analysis and machine intelligence},
  2020.

\bibitem{chen2017rethinking}
Liang-Chieh Chen, George Papandreou, Florian Schroff, and Hartwig Adam,
\newblock ``Rethinking atrous convolution for semantic image segmentation,''
\newblock {\em arXiv preprint arXiv:1706.05587}, 2017.

\end{thebibliography}

\end{document}